\documentclass[conference]{IEEEtran}
\IEEEoverridecommandlockouts
\usepackage{cite}
\usepackage{amsmath,amssymb,amsfonts}
\usepackage{algorithmic}
\usepackage{graphicx}
\usepackage{float}
\usepackage{textcomp}
\usepackage{xcolor}
\def\BibTeX{{\rm B\kern-.05em{\sc i\kern-.025em b}\kern-.08em
    T\kern-.1667em\lower.7ex\hbox{E}\kern-.125emX}}
\begin{document}

\title{Optimizing Task Scheduling in Heterogeneous Computing Environments: A Comparative Analysis of CPU, GPU, and ASIC Platforms Using E2C Simulator \\
}

\author{\IEEEauthorblockN{1\textsuperscript{st} Ali Mohammadjafari, 1\textsuperscript{st} Poorya Khajouie,}
 \\

}

\maketitle

\begin{abstract}
Efficient task scheduling in heterogeneous computing environments is imperative for optimizing resource utilization and minimizing task completion times. In this study, we conducted a comprehensive benchmarking analysis to evaluate the performance of four scheduling algorithms—First-Come, First-Served (FCFS), FCFS with No Queuing (FCFS-NQ), Minimum Expected Completion Time (MECT), and Minimum Expected Execution Time (MEET)—across varying workload scenarios. We defined three workload scenarios: low, medium, and high, each representing different levels of computational demands. Through rigorous experimentation and analysis, we assessed the effectiveness of each algorithm in terms of total completion percentage, energy consumption, wasted energy, and energy per completion. Our findings highlight the strengths and limitations of each algorithm, with MECT and MEET emerging as robust contenders, dynamically prioritizing tasks based on comprehensive estimates of completion and execution times. Furthermore, MECT and MEET exhibit superior energy efficiency compared to FCFS and FCFS-NQ, underscoring their suitability for resource-constrained environments. This study provides valuable insights into the efficacy of task scheduling algorithms in heterogeneous computing environments, enabling informed decision-making to enhance resource allocation, minimize task completion times, and improve energy efficiency

\end{abstract}


\section{Introduction}
Process scheduling is a fundamental aspect of modern computing systems, that is playing a critical role in determining the efficiency and responsiveness of  computer operations. At the main part of this mechanism lies the ability to wisely allocate CPU time to multi-processes. The importance of scheduling cannot be overstated; it is the arbiter of resource distribution, ensuring that each process receives a fair share of computing power while minimizing idle time. This delicate balance is crucial for maintaining system stability and achieving optimal performance.

The historical development of process scheduling in computer systems is back to several decades, reflecting the evolution of computing technologies and the increasing demand for efficient resource utilization in computer technology. In the early days of computing, scheduling was relatively simplistic, with batch processing systems executing tasks in a sequential manner. However, as computing environments became more complex and diverse, the need for more sophisticated scheduling algorithms became a critical problem. One of the earliest milestones in process scheduling was the introduction of the first multitasking operating systems, such as IBM's OS/360 in the 1960s, which allowed multiple programs to run concurrently on a single processor. This innovation built the way for the development of modern scheduling algorithms, with seminal works like Edsger Dijkstra's "The Structure of the 'THE'-Multiprogramming System" in 1968, which introduced concepts such as process states and priority scheduling. In the recent decades, researchers have continued to refine and innovate upon scheduling algorithms, incorporating advancements in computer architecture, real-time computing, and parallel processing. So, Finding the optimal strategy for workload scheduling is still a problem\cite{fiad2020improved}. It involves a deep understanding of the intricacies of both the hardware and the software it supports. By adapting the scheduling algorithm to the specific characteristics of the workload and the existing hardware, one can significantly enhance system throughput. This not only reduces waiting times for processes but also maximizes the overall efficiency of the system, paving the way for more advanced and responsive operating systems in our increasingly digital world.

\subsection{Scheduling}
Job scheduling in CPU context refers to the process of allocating computing resources to execute a set of jobs, where each job has specific computational requirements and resource demands. The goal is to optimize the use of CPU resources, improve system performance, and ensure that all jobs are completed in an efficient manner\cite{7524417}. In Figure 1, we try to show different type of CPU scheduling algorithms.

Scheduling strategies are categorized in two main groups:
\begin{itemize}
    \item Non-preemptive Scheduling: Non-preemptive Scheduling: This strategy involves scheduling jobs so that once a job starts executing, it runs to completion without being interrupted. This is essential in systems where interrupting a job can be costly or is not allowed due to system constraints.
    \item Priority-based Algorithms: These algorithms prioritize jobs based on certain criteria such as job size, demand, or a combination of both (volume). Examples include Shortest Job First (SJF), Smallest Demand First (SDF), and Smallest Volume First (SVF)
\end{itemize}


   \begin{figure}
    \centering
    \includegraphics[width=.4\textwidth]{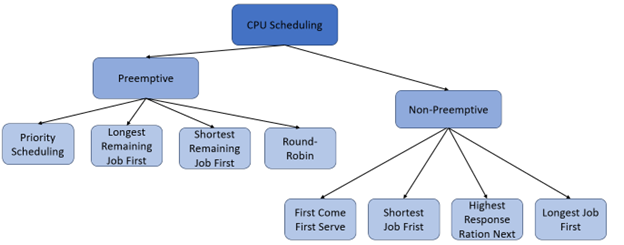}
    \caption{Different types of CPU Scheduling Algorithms}
\label{fig:high_level}
\vspace{-.3 cm}
\end{figure}

\begin{table*}[t]
\centering
\caption{Comparison of CPU Scheduling Algorithms}
\resizebox{\textwidth}{!}{%
\begin{tabular}{|c|c|c|c|c|c|c|}
\hline
Algorithm &
  Allocation   is &
  Complexity &
  AWT &
  Preemption &
  Starvation &
  Performance \\ \hline
FCFS &
  According to   arrival time, CPU is allocated. &
  Simple and   easy to implement &
  Large &
  No &
  No &
  Slow   performance \\ \hline
SJF &
  Based on lowest CPU burst time (BT). &
  More complex   than FCFS &
  Smaller than   FCFS &
  No &
  Yes &
  Minimum   AWT \\ \hline
LJFS &
  Based on highest CPU burst time (BT) &
  More complex   than FCFS &
  Depending on   some measures &
  No &
  Yes &
  Big   turn-around time \\ \hline
LRTF &
  Same as LJFS   but preemptive &
  More complex   than FCFS &
  Depending   on some measures &
  Yes &
  Yes &
  Preference   to longer jobs \\ \hline
SRTF &
  Same as SJF   but preemptive &
  More complex   than FCFS &
  Depending on   some measures &
  Yes &
  Yes &
  Preference   to short jobs \\ \hline
RR &
  According to   order of process with fixed TQ &
  Complexity   depends on TQ size &
  Large as   compared to SJF and Priority &
  Yes &
  No &
  Each   process has fixed time \\ \hline
Priority Pre. &
  According to   priority. Bigger priority task executes first &
  Less complex &
  Smaller than   FCFS &
  Yes &
  Yes &
  Well   performance but contains starvation \\ \hline
Priority Non-pre. &
  According to   priority with monitoring incoming higher priority jobs &
  Less complex than Priority Pre. &
  Smaller than   FCFS &
  No &
  Yes &
  Beneficial   with batch systems \\ \hline
MLQ &
  According to   process in bigger queue priority &
  More complex   than priority algorithms &
  Smaller than   FCFS &
  No &
  Yes &
  Good   performance but contains starvation \\ \hline
MFLQ &
  According to   process of bigger priority queue &
  Most   complex depends on TQ size &
  Smaller than   other types &
  No &
  No &
  Good   performance \\ \hline
\end{tabular}%
}
\end{table*}

As we said before, in the realm of computer systems and operating environments, optimizing process scheduling is a critical goal to enhance system performance and resource utilization. The selection of an appropriate CPU scheduling algorithm profoundly influences various system metrics, including throughput, response time, and resource utilization efficiency\cite{harki2020cpu}. Consequently, comparing CPU scheduling algorithms serves as a fundamental aspect of system optimization, enabling researchers and practitioners to evaluate and understand the strengths and weaknesses of different scheduling strategies under various workload scenarios. By conducting such comparisons, insights are gained into how each algorithm performs under different conditions, aiding in the selection or development of algorithms tailored to specific system requirements. In table1 we can see a detailed comparison of CPU scheduling algorithms. This comparative analysis not only facilitates the identification of optimal scheduling strategies but also contributes to the advancement of theoretical knowledge and practical implementations in the field of process scheduling, ultimately leading to more efficient and responsive computing systems\cite{omar2021comparative}.

\subsection{Understanding the Significance of Process Scheduling Optimization}

In today's fast-changing computing landscape, where we can see diverse environments such as cloud computing, real-time systems, and multi-core processors, the imperative for efficient process scheduling has never been more pronounced. The escalating complexity and heterogeneity of these environments underscore the critical role of scheduling algorithms in achieving optimal system performance. Efficient process scheduling is not just merely a technical concern; it directly impacts user experience, application responsiveness, and overall system efficiency\cite{gupta2022efficient}. Delays in process execution can lead to poor performance and reduced user satisfaction, highlighting the profound influence of scheduling decisions on the end-user experience. However, creating effective scheduling algorithms needs to answer to different challenges and trade-offs. Balancing considerations such as fairness, throughput, latency, and resource utilization against real-world constraints and system dynamics is a daunting task faced by system designers and researchers alike. Despite these challenges, process scheduling remains integral to a diverse array of applications and domains, spanning batch processing, scientific computing, multimedia applications, and embedded systems\cite{abualigah2021novel}. The evaluation of scheduling algorithms relies on a variety of metrics, including average waiting time, turnaround time, response time, fairness, and scalability, each of which plays a crucial role in assessing algorithm effectiveness and guiding further optimization efforts. 
\subsection{E2C: A Visual Simulator for Heterogeneous Computing Systems }

The E2C simulator, introduced in \cite{mokhtari2023e2c}, is a cutting-edge visual tool designed to address the complexities of heterogeneous computing systems. It offers a cost-effective and time-efficient solution for studying the performance of diverse system configurations, a challenge that has long plagued researchers and practitioners in the field of distributed computing. By enabling users to simulate heterogeneous computing systems, implement scheduling methods, measure energy consumption, and visualize system performance, E2C opens up a realm of possibilities for students, researchers, and industry professionals alike. This simulator not only provides a practical platform for exploring system heterogeneity but also serves as a valuable resource for understanding the intricacies of dynamic resource allocation, the concept of dark silicon, specialized data centers for planet-scale applications, and the multi-tenancy of latency-sensitive deep learning applications on the edge. With its comprehensive features, E2C stands as a pivotal tool in advancing the study and application of heterogeneous computing systems, offering insights and opportunities for innovation in this rapidly evolving field. So in this paper we used this simulator to generate workloads in different scenario, then we used these scenarios to do a computational task scheduling benchmark.

\subsection{Proposed Approach and Research Framework}

In this paper, we present a comprehensive analysis of task scheduling across heterogeneous computing environments, including CPU, GPU, and ASIC. Our research framework aims to evaluate the efficacy of various scheduling algorithms in managing workloads of varying volumes, categorized as low, medium, and high.

We delineate three distinct scenarios based on workload volume to encapsulate the diverse operational conditions encountered in real-world computing environments. These scenarios provide a structured framework for assessing the performance of scheduling algorithms under different levels of computational demand.

We employ four distinct scheduling algorithms to orchestrate task execution within each scenario, each tailored to address specific optimization objectives:

\begin{itemize}
    \item First-Come, First-Served (FCFS): A rudimentary scheduling approach where tasks are executed in the order of their arrival, irrespective of their computational requirements or expected completion times.
    \item FCFS with No Queuing (FCFS-NQ): A variant of FCFS that eliminates queuing, thereby prioritizing immediate task execution over waiting, which may be beneficial in scenarios where latency is critical.
    \item Minimum Expected Completion Time (MECT): This algorithm prioritizes tasks based on their expected completion times, aiming to minimize the overall turnaround time of the task queue.
    \item Minimum Expected Execution Time (MEET): MEET prioritizes tasks based on their expected execution times, focusing on minimizing resource utilization and maximizing throughput efficiency.
\end{itemize}

Our research framework facilitates a comparative analysis of the performance of each scheduling algorithm across the defined scenarios. By evaluating metrics such as throughput, latency, and resource utilization, we aim to discern the strengths and weaknesses of each algorithm in diverse operational contexts. Furthermore, we seek to identify the optimal scheduling strategy for each workload scenario, providing insights for efficient resource management in heterogeneous computing environments.

This study contributes to the field of task scheduling in heterogeneous computing environments by offering a systematic evaluation of scheduling algorithms under various workload scenarios. Our findings shed light on the performance characteristics of different scheduling strategies, providing valuable insights for system designers and administrators. Additionally, our research addresses the practical challenge of optimizing resource utilization in heterogeneous computing environments, thereby enhancing the overall efficiency and performance of computational systems. By finding the strengths and weaknesses of each scheduling algorithm across different workload scenarios, our work serv`es as a valuable reference for guiding decision-making processes in real-world computing environments.

The rest of this paper is organized as follows. Section II discusses background study and related prior works. We explain our methodology and our benchmark in detail in Section III. Next, we discuss experimental evaluation and performance analysis in Section IV. Finally, Section V concludes the paper and provides a few avenues for the future studies.

\section{Related Works}

\subsection{Benchmarking Task Scheduling Algorithms}
As we delve into the realm of heterogeneous computing systems, the efficacy of task scheduling algorithms becomes paramount. Maurya and Tripathi1 in 2018 presents a benchmarking approach to compare the performance of various scheduling algorithms in a heterogeneous computing environment. The authors employ a comprehensive set of metrics to assess the efficiency, effectiveness, and scalability of the algorithms \cite{maurya2018benchmarking}. In this concept, \cite{madhura2023efficient} \cite{thonglek2023benchmarks} \cite{carracciuolo2023toward} Also have similar research on doing benchmark on heterogeneous computing systems. 
Sirisah and Prasad in 2023 proposed the novel MPEFT (Maximizing Parallelism for Minimizing Earliest Finish Time) algorithm, which aims to minimize makespan by enhancing parallelism within workflows. Through rigorous evaluation against classical scheduling algorithms, the MPEFT algorithm demonstrates superior performance in terms of makespan, speedup, efficiency, and frequency of best results. The study contributes valuable insights to the field of high-performance computing, emphasizing the importance of effective task scheduling in maximizing system performance \cite{sirisha2023mpeft}.

Sirsha in 2023 presented two novel heuristic scheduling algorithms, Global Highest degree Task First (GHTF) and Critical Path/Earliest Finish Time (CP/EFT), aimed at minimizing the schedule length in Heterogeneous Computing Environments. Both algorithms show significant improvements over classical list scheduling algorithms, with GHTF achieving better schedules by 5–20\% and CP/EFT by 10.76–23.45\%3. The paper also includes a comprehensive survey of existing algorithms, detailed descriptions of the proposed algorithms, and an evaluation of their performance against widely referred scheduling algorithms \cite{sirisha2023complexity}.

\subsection{Machine Learning Approaches:}

In recent years, researchers have been exploring the integration of machine learning techniques with process scheduling to enhance system efficiency. By leveraging approaches like reinforcement learning, supervised learning, and unsupervised learning, these methods aim to use historical workload data to make smarter scheduling decisions. This subsection examines the literature on machine learning-based scheduling algorithms and assesses their effectiveness in improving system performance.
In a recent years a lot of studies provide a comprehensive overview of machine learning-based scheduling research, highlighting the growth of this field and the potential for future advancements. they discuss various machine learning algorithms and scheduling scenarios, emphasizing the importance of deep learning and reinforcement learning-based algorithms\cite{li2021machine} \cite{song2023reinforcement} \cite{liu2023heterps} \cite{cheng2023deep}. In 2023 Mangalampalli et. al. discusse the use of Deep Reinforcement Learning-based approaches for optimal task mapping in cloud computing, focusing on improving metrics such as makespan, energy consumption, and SLA violation. They compare the proposed algorithm with existing mechanisms and demonstrates superior performance. The study also emphasizes the importance of dynamic task scheduling based on priorities to minimize inefficiencies in existing scheduling mechanisms.\cite{mangalampalli2024drlbtsa}. 

In 2023, Hayat et. al. proposed a solution that combines a load-balanced task scheduler with a machine learning-based device predictor. By forecasting execution times on CPU and GPU devices and assigning tasks accordingly, the proposed approach aims to optimize system performance. To counteract the risk of tasks predominantly mapping to a single device, a work-stealing-based task scheduler dynamically redistributes workload among devices\cite{hayat2023machine}. 

\subsection{Adaptive Scheduling Strategies:}

In the realm of process scheduling optimization, adaptive strategies take center stage, dynamically adjusting scheduling decisions to match real-time workload variations. These strategies embody responsiveness, employing techniques such as feedback control, dynamic priority adjustment, and online learning algorithms to efficiently manage fluctuating workloads. As system dynamics grow more intricate, exploring adaptive scheduling becomes essential, offering improved performance and efficiency in task management. This subsection delves into the domain of adaptive scheduling strategies, unveiling innovative approaches that enhance process scheduling adaptability and robustness. Mao et. al. in a paper propose a novel approach by decomposing scheduling strategies into three dimensions and making informed decisions along each dimension based on workload characteristics. Compared to existing approaches, MorphStream achieves significantly higher throughput, up to 3.4 times, and reduced processing latency by 69.1\% when handling real-world scenarios characterized by complex and dynamically changing workload dependencies\cite{mao2023morphstream}. Zheng er. al. propose Shockwave, a forward-thinking scheduler. Drawing upon principles from market theory and stochastic dynamic programming, Shockwave aims to optimize both efficiency and fairness in dynamically changing environments. Experimental results indicate that Shockwave significantly outperforms existing fair scheduling schemes, enhancing makespan by 1.3× and fairness by 2× for ML job traces characterized by dynamic adaptation\cite{zheng2023shockwave}.  Li et. al introduce an adaptive batch-stream scheduling method, a chiplet-based core-cluster binding mechanism, and a chiplet-based nearest task stealing approach to enhance computing efficiency\cite{cai2024abss}.

\subsection{Real-Time Scheduling}

In the realm of CPU scheduling, real-time scheduling algorithms hold a critical role in ensuring timely execution of tasks with stringent timing constraints. These algorithms are specifically designed to meet the requirements of real-time systems where tasks must complete within predetermined deadlines to guarantee system stability and reliability. Real-time scheduling for CPUs encompasses various approaches, such as Rate-Monotonic Scheduling (RMS), Earliest Deadline First (EDF), and Deadline-Monotonic Scheduling (DMS), each offering unique strategies for task prioritization based on deadlines or rates\cite{cai2024abss}. Saifullah et. al introduce techniques to minimize CPU energy consumption while ensuring that all tasks meet their deadlines. They propose a novel integration of a frequency optimization engine and dynamic voltage and frequency scaling (DVFS) with classical real-time scheduling policies\cite{saifullah2020cpu}. In 2021 Kang et. al introduce a novel approach that combines CPU-friendly quantization with fine-grained CPU/GPU allocation, aiming to improve the execution of DNN tasks while preserving timing guarantees and minimizing accuracy loss\cite{kang2021lalarand}. Yoon et. al. introduce a novel swap scheme aimed at reducing memory power consumption by leveraging high-speed NVM storage and optimizing power savings in both CPU and memory\cite{yoon2022supporting}. 

\section{Methodology}
Before delving into the intricacies of our methodology, it's essential to underscore the significance of task scheduling in heterogeneous computing environments. As computing systems evolve, the integration of diverse processing units, such as CPUs, GPUs, and ASICs, has become increasingly prevalent, offering unprecedented computational capabilities\cite{mokhtari2022felare}. However, harnessing the full potential of these heterogeneous architectures necessitates efficient task scheduling strategies tailored to their unique characteristics and operational requirements.

In this study, we embark on a comprehensive exploration of task scheduling across CPU, GPU, and ASIC platforms. Our methodology revolves around the systematic evaluation of scheduling algorithms under varying workload conditions, categorized into low, medium, and high volumes by using E2C Simulator \cite{mokhtari2023e2c}. By leveraging a diverse set of scheduling algorithms—First-Come, First-Served (FCFS), FCFS with No Queuing (FCFS-NQ), Minimum Expected Completion Time (MECT), and Minimum Expected Execution Time (MEET)—we aim to discern the optimal scheduling strategy for each workload scenario.

\subsection{E2C Simulator}

In our study, we used E2C Simulator that is published in \cite{mokhtari2023e2c}. E2C developed by external researchers, to evaluate the performance of various scheduling algorithms across different workload scenarios. The simulator served as a crucial tool in our investigation, allowing us to model and analyze task execution behavior on CPU, GPU, and ASIC platforms. In figure \ref{fig:overview}, an overview of the E2C Simulator is shown. E2C core is available for download at the following address: https://github.com/hpcclab/E2C-Sim The manual document on how to run E2C and its options and full documentations are available here: https://hpcclab.github.io/E2C-Sim-docs/

   \begin{figure}[h]
    \centering
    \includegraphics[width=.45\textwidth]{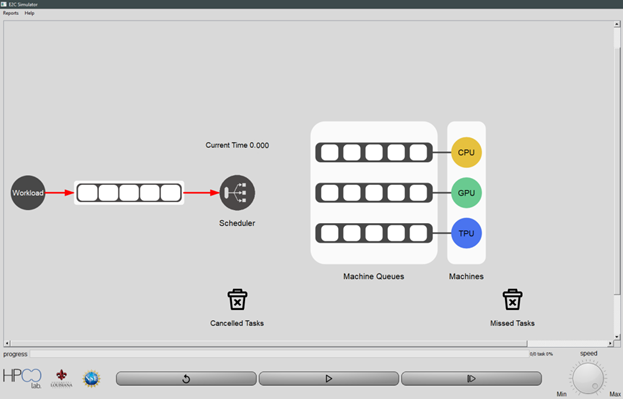}
    \caption{Summary of the E2C Simulator comprising essential elements, namely the input workload, a queue for incoming tasks, a scheduler (also known as a load balancer), and a variety of machines depicted in distinct hues \cite{mokhtari2023e2c}.}
\label{fig:overview}
\vspace{-.3 cm}
\end{figure}
\subsubsection{Simulation Components and Features}

\begin{table*}[]
\centering
\caption{Overview of Scenarios }
\begin{tabular}{cllllll}
\cline{1-6}
\multicolumn{1}{|c|}{} &
  \multicolumn{1}{c|}{\# of Task1} &
  \multicolumn{1}{c|}{\# of Task2} &
  \multicolumn{1}{c|}{\# of Task3} &
  \multicolumn{1}{c|}{Start time} &
  \multicolumn{1}{c|}{End time} &
  \multicolumn{1}{c}{} \\ \cline{1-6}
\multicolumn{1}{|c|}{Scenario1: Low Workload} &
  \multicolumn{1}{c|}{\begin{tabular}[c]{@{}c@{}}700\\ Normal distribution\end{tabular}} &
  \multicolumn{1}{c|}{\begin{tabular}[c]{@{}c@{}}700\\ Exponential distribution\end{tabular}} &
  \multicolumn{1}{c|}{\begin{tabular}[c]{@{}c@{}}700\\ Uniform distribution\end{tabular}} &
  \multicolumn{1}{c|}{0} &
  \multicolumn{1}{c|}{1000 ms} &
  \multicolumn{1}{c}{} \\ \cline{1-6}
\multicolumn{1}{|c|}{Scenario2: Medium Workload} &
  \multicolumn{1}{c|}{\begin{tabular}[c]{@{}c@{}}1000\\ Normal distribution\end{tabular}} &
  \multicolumn{1}{c|}{\begin{tabular}[c]{@{}c@{}}1000\\ Exponential distribution\end{tabular}} &
  \multicolumn{1}{c|}{\begin{tabular}[c]{@{}c@{}}1000\\ Uniform distribution\end{tabular}} &
  \multicolumn{1}{c|}{0} &
  \multicolumn{1}{c|}{1000ms} &
  \multicolumn{1}{c}{} \\ \cline{1-6}
\multicolumn{1}{|c|}{Scenario3: High Workload} &
  \multicolumn{1}{c|}{\begin{tabular}[c]{@{}c@{}}1400\\ Normal distribution\end{tabular}} &
  \multicolumn{1}{c|}{\begin{tabular}[c]{@{}c@{}}1400\\ Exponential distribution\end{tabular}} &
  \multicolumn{1}{c|}{\begin{tabular}[c]{@{}c@{}}1400\\ Uniform distribution\end{tabular}} &
  \multicolumn{1}{c|}{0} &
  \multicolumn{1}{c|}{1000ms} &
  \multicolumn{1}{c}{} \\ \cline{1-6}
 & \multicolumn{1}{c}{} & \multicolumn{1}{c}{} & \multicolumn{1}{c}{} & \multicolumn{1}{c}{} & \multicolumn{1}{c}{} & \multicolumn{1}{c}{} \\        
\end{tabular}
\label{scenarios}
\end{table*}

\begin{itemize}
    \item Task Generation:
        The simulator enabled us to generate synthetic workloads representing low, medium, and high task volumes. These workloads consisted of tasks with varying execution times, resource requirements, and dependencies.
        Customization options allowed us to define the number of tasks, their characteristics, and the inter-task relationships.
    \item Resource Models: Detailed resource models were incorporated         for CPU, GPU, and ASIC devices. These models considered factors such as processing speed, memory capacity, and communication latency.
    
          By accurately simulating resource availability and contention during task execution, we ensured realistic scenarios
    \item Scheduling Algorithms:
          Within the E2C Simulator, we implemented four scheduling algorithms:
          \begin{itemize}
              \item First-Come-First-Serve (FCFS): Tasks executed in the order of arrival.
              \item FCFS with No Queuing (FCFS-NQ): Similar to FCFS, but tasks could not wait in a queue
              \item Minimum Expected Completion Time (MECT): Prioritized tasks with the shortest expected completion time.
              \item Minimum Expected Execution Time (MEET): Prioritized tasks with the shortest expected execution time.
          \end{itemize}
         The simulator allowed us to switch between these algorithms dynamically, observing their impact on overall task completion times.
\end{itemize}
The E2C Simulator emerged as a valuable asset, enabling us to evaluate scheduling algorithms across diverse workload scenarios. Its flexibility allowed us to explore trade-offs and make informed recommendations for practical task scheduling in heterogeneous computing environments\cite{mokhtari2023e2c}.

\subsection{Workload Generation and Task Specifications}

In our study, the generation of workloads and the specification of task characteristics play a crucial role in accurately simulating real-world scenarios and evaluating the efficacy of different scheduling algorithms across CPU, GPU, and ASIC platforms. We meticulously design our workload generation process to encompass a wide range of computational demands and timing constraints.

Our workload generation process begins with the classification of tasks into three distinct types, each carefully crafted to represent varying degrees of computational intensity and timing sensitivity:

\begin{itemize}
    \item Task1: This task type is characterized by a mean data size of 100 KB and a slack of 2 milliseconds. Task1 represents computational tasks with moderate data requirements and relatively lenient timing constraints.
    \item Task2: With a mean data size of 75 KB and a slack of 1.5 milliseconds, Task2 embodies computational tasks with slightly lower data demands but more stringent timing requirements compared to Task1.
    \item Task3: Task3, with a mean data size of 50 KB and a slack of 1 millisecond, represents computational tasks with relatively low data requirements but tight timing constraints, requiring swift execution.
\end{itemize}

These task types are carefully selected to cover a spectrum of computational requirements commonly encountered in practical applications, ensuring a comprehensive evaluation of scheduling algorithms.

\begin{table}[h]
\centering
\caption{Machine Types and Task Execution Time (EET) on Respective Platforms}
\begin{tabular}{cllllll}
\hline
\multicolumn{1}{|c|}{} &
  \multicolumn{3}{c|}{EET( Expected Execution Time)} &
  \multicolumn{1}{c|}{Power} &
  \multicolumn{1}{c|}{Idle Power} &
  \multicolumn{1}{c|}{\#Replicas} \\ \cline{1-4}
\multicolumn{1}{|c|}{} &
  \multicolumn{1}{c|}{Task1} &
  \multicolumn{1}{c|}{Task2} &
  \multicolumn{1}{c|}{Task3} &
  \multicolumn{1}{c|}{} &
  \multicolumn{1}{c|}{} &
  \multicolumn{1}{c|}{} \\ \hline
\multicolumn{1}{|c|}{CPU} &
  \multicolumn{1}{c|}{5ms} &
  \multicolumn{1}{c|}{4ms} &
  \multicolumn{1}{c|}{3ms} &
  \multicolumn{1}{c|}{150} &
  \multicolumn{1}{c|}{15} &
  \multicolumn{1}{c|}{2} \\ \hline
\multicolumn{1}{|c|}{GPU} &
  \multicolumn{1}{c|}{2ms} &
  \multicolumn{1}{c|}{1.5ms} &
  \multicolumn{1}{c|}{1ms} &
  \multicolumn{1}{c|}{300} &
  \multicolumn{1}{c|}{30} &
  \multicolumn{1}{c|}{4} \\ \hline
\multicolumn{1}{|c|}{ASIC} &
  \multicolumn{1}{c|}{1ms} &
  \multicolumn{1}{c|}{0.8ms} &
  \multicolumn{1}{c|}{0.5ms} &
  \multicolumn{1}{c|}{50} &
  \multicolumn{1}{c|}{5} &
  \multicolumn{1}{c|}{2} \\ \hline

\end{tabular}
\label{mt}
\end{table}

To accurately model task execution across different computing platforms, we estimate the execution time (EET) for each task type on CPU, GPU, and ASIC platforms. Table \ref{mt} shows all this information. 

In table \ref{scenarios}, to comprehensively evaluate the performance of scheduling algorithms under various workload conditions, we define three distinct scenarios: low, medium, and high, each with specific task counts and arrival distributions. 

\begin{table}[h]
\centering
\caption{Sample of Low Scenario Workload}
\begin{tabular}{|l|l|l|l|lll}
\cline{1-4}
task\_type & data\_size & arrival\_time & deadline & \multicolumn{1}{c}{} & \multicolumn{1}{c}{} & \multicolumn{1}{c}{} \\ \cline{1-4}
Task1 & 110.899 & 692.529 & 694.529 & \multicolumn{1}{c}{} & \multicolumn{1}{c}{} & \multicolumn{1}{c}{} \\ \cline{1-4}
Task1 & 110.663 & 668.942 & 670.942 & \multicolumn{1}{c}{} & \multicolumn{1}{c}{} & \multicolumn{1}{c}{} \\ \cline{1-4}
Task1 & 115.118 & 306.436 & 308.436 & \multicolumn{1}{c}{} & \multicolumn{1}{c}{} & \multicolumn{1}{c}{} \\ \cline{1-4}
Task1 & 100.568 & 617.725 & 619.725 & \multicolumn{1}{c}{} & \multicolumn{1}{c}{} & \multicolumn{1}{c}{} \\ \cline{1-4}
Task2 & 70.814  & 5.612   & 7.112   &                      &                      &                      \\ \cline{1-4}
Task2 & 45.237  & 7.775   & 9.275   &                      &                      &                      \\ \cline{1-4}
Task2 & 77.869  & 10.118  & 11.618  &                      &                      &                      \\ \cline{1-4}
Task2 & 72.863  & 10.903  & 12.403  &                      &                      &                      \\ \cline{1-4}
Task3 & 38.241  & 624.736 & 625.736 &                      &                      &                      \\ \cline{1-4}
Task3 & 43.915  & 493.478 & 494.478 &                      &                      &                      \\ \cline{1-4}
Task3 & 52.198  & 159.911 & 160.911 &                      &                      &                      \\ \cline{1-4}
Task3 & 47.776  & 709.633 & 710.633 &                      &                      &                      \\ \cline{1-4}
\end{tabular}
\end{table}

This meticulous approach to workload generation and scenario definition ensures a comprehensive evaluation of scheduling algorithms under diverse workload conditions, facilitating informed decision-making regarding task allocation and optimization strategies across different computing platforms.

\subsection{Scheduling Algorithms}
In this part, we provide a comprehensive overview of the scheduling algorithms utilized in the study and elaborate on the methodology employed for workload simulation and algorithm selection.
\begin{itemize}
    \item First-Come, First-Served (FCFS): FCFS is a classic scheduling algorithm that prioritizes tasks based on their arrival time. Tasks are executed in the order they enter the system, without consideration for their urgency or execution time.
    \item FCFS with No Queuing (FCFS-NQ): FCFS-NQ is an extension of FCFS designed to mitigate queuing delays. It rejects incoming tasks when the system is at full capacity, thereby preventing queuing and ensuring immediate task rejection if resources are unavailable.
    \item Minimum Expected Completion Time (MECT): MECT is a dynamic scheduling algorithm that estimates the expected completion time for each task based on factors such as urgency, deadlines, and resource availability. It prioritizes tasks to minimize the overall completion time of the workload.
    \item Minimum Expected Execution Time (MEET): MEET predicts the execution time for each task and schedules them to minimize the total expected execution time. It considers task characteristics such as complexity and resource requirements to optimize task execution across heterogeneous architectures.
\end{itemize}

\subsection{Workload Simulation Methodology}

In order to assess the efficacy of scheduling algorithms across diverse computational architectures, a meticulous methodology was devised to simulate workloads and evaluate algorithmic performance. This methodology encompasses several key stages, each contributing to the comprehensive analysis of scheduling strategies. 

\subsubsection{Configuration Setup}
A robust configuration setup was paramount to ensuring the fidelity of the simulation environment. To this end, meticulous attention was paid to specifying configuration parameters within the  \emph{config.json} file. These parameters encapsulated crucial aspects such as machine specifications, task types, and battery capacity. Notably, machine types including CPU, GPU, and ASIC were defined with their respective power consumption and number of replicas, thereby laying the foundation for a heterogeneous computational environment.

\subsubsection{Algorithm Selection and Execution}

Workloads are generated using CSV files, where each file contains task arrival times and associated parameters. The workload generator ensures variability in workload intensity and composition to simulate real-world scenarios accurately. 
\begin{verbatim}
# Example code for generating workloads
path_to_arrivals = f'./workloads/{scenario}/
    workload-0.csv'
\end{verbatim}

The simulation framework allows for the selection of scheduling algorithms through the \emph{config.json} file. By modifying the  \emph{"scheduling-method"} parameter, different algorithms can be chosen for evaluation.

\begin{verbatim}
#Example code for selecting scheduling 
#algorithm in config.json
"scheduling_method": "FCFS"

\end{verbatim}

The simulation engine processes the generated workloads using the selected scheduling algorithm. Performance metrics such as completion percentage, energy consumption, and wasted energy are recorded and analyzed for each workload scenario.

\begin{verbatim}
# Example code for running simulation and 
#recording results

simulation = Simulator(path_to_arrivals,
    path_to_etc, path_to_reports, seed=123)
simulation.set_scheduling_method
    (config.scheduling_method)
simulation.run()

\end{verbatim}

By following these steps, researchers can explore the impact of different scheduling algorithms on workload management across diverse computational architectures, facilitating informed decision-making in resource allocation and task prioritization.

\section{Evaluation of Task Scheduling Algorithms in Heterogeneous Computing Environments}

Task scheduling in heterogeneous computing environments is a multifaceted endeavor, necessitating the selection of optimal algorithms to ensure efficient resource utilization and task completion. In this study, we conducted a comprehensive benchmarking analysis, evaluating four distinct scheduling algorithms—First-Come, First-Served (FCFS), FCFS with No Queuing (FCFS-NQ), Minimum Expected Completion Time (MECT), and Minimum Expected Execution Time (MEET)—across three scenarios of varying workload volumes: low, medium, and high. Through rigorous experimentation and analysis, we aimed to discern the strengths and weaknesses of each algorithm and elucidate their suitability for different workload scenarios.

\subsection{Effectiveness Across Workload Scenarios:}

\begin{figure}[h]
    \centering
    \includegraphics[width=.5\textwidth]{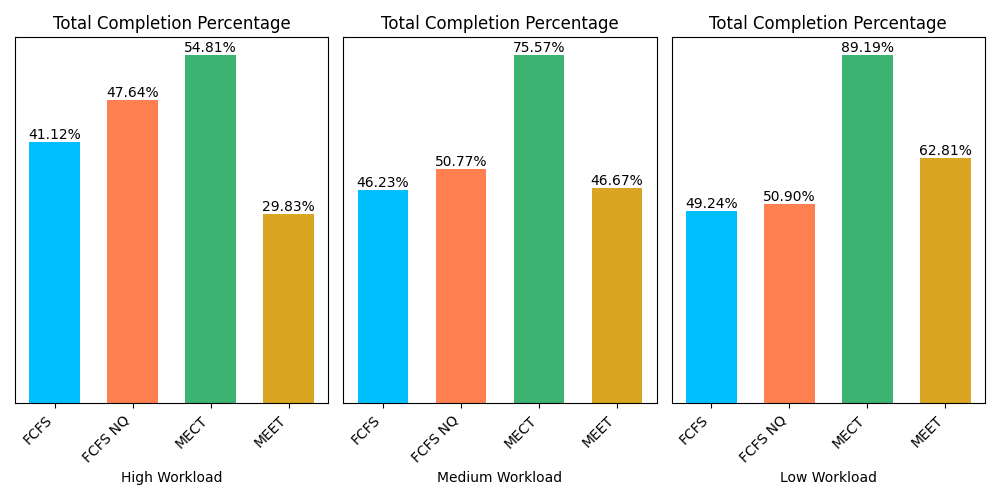}
    \caption{Total Completion Percentage.}
\label{fig:TCP}
\vspace{-.3 cm}
\end{figure}

In our evaluation across varied workload scenarios, each scheduling algorithm exhibited distinct performance attributes, shedding light on their efficacy under different computational demands. In scenarios characterized by low workload volumes, FCFS and FCFS-NQ showcased commendable performance, leveraging their straightforward execution strategy to efficiently manage task allocation. However, as workload volumes escalated, particularly in scenarios with stringent task deadlines and varying urgency levels, the limitations of these algorithms became apparent.As it is shown in figure \ref{fig:TCP}, the total completion percentage for FCFS and FCFS-NQ dwindled notably as workload intensity increased, reaching 49.24\% and 50.9\% respectively in low workload scenarios, 46.23\% and 50.77\% in medium workload scenarios, and 41.12\% and 47.64\% in high workload scenarios.

In stark contrast, MECT and MEET emerged as formidable contenders across all workload scenarios, adeptly adapting to fluctuating computational demands. By dynamically prioritizing tasks based on comprehensive completion and execution time estimates, these algorithms optimized resource utilization and minimized task completion times. In low workload scenarios, MECT and MEET demonstrated substantial improvements in total completion percentage, achieving 89.19\% and 62.81\% respectively, showcasing their ability to efficiently manage tasks even in scenarios with relatively modest computational demands. As workload volumes increased to medium and high levels, the resilience of MECT became even more pronounced, with total completion percentages of 75.57\% in medium workload scenarios, and 54.81\% in high workload scenarios, respectively.

In high and medium workload scenarios, the computational resources may become saturated due to the increased volume of tasks competing for execution. MEET, which prioritizes tasks based on expected execution time, may struggle to efficiently allocate resources when the demand exceeds capacity. This can lead to increased contention and resource bottlenecks, ultimately impacting task completion rates.

These findings underscore the critical importance of selecting appropriate scheduling algorithms tailored to the specific characteristics of the workload. While FCFS-based algorithms offer simplicity and ease of implementation, MECT emerges as indispensable tools for scenarios necessitating optimized resource utilization and efficient task completion, particularly in high-pressure environments with stringent deadlines and fluctuating computational demands.

\subsection{Resource Utilization and Energy Efficiency:}

\subsubsection{Effectiveness Across Workload Scenarios}

In our assessment of scheduling algorithms across diverse workload scenarios, each algorithm exhibited distinctive performance attributes, showcasing their inherent strengths and limitations. Particularly in low workload scenarios, FCFS and FCFS-NQ demonstrated commendable efficiency, leveraging their simplicity to manage task execution effectively. However, as workload volumes escalated, particularly in scenarios with stringent task deadlines and varying urgency levels, the limitations of these algorithms became evident. In contrast, MECT and MEET emerged as robust contenders, dynamically prioritizing tasks based on completion and execution time estimates, thereby optimizing resource utilization and minimizing task completion times.

\subsubsection{Energy Consumption Analysis}

A crucial aspect of algorithm evaluation pertains to energy consumption, a metric essential for assessing the sustainability and efficiency of task scheduling strategies. Across all workload scenarios, MECT and MEET exhibited superior energy efficiency compared to FCFS and FCFS-NQ. In figure \ref{fig:TCE}, in low workload scenarios, MECT and MEET demonstrated a marked reduction in total consumed energy, with values of 22.73\% and 16.14\% respectively, compared to FCFS and FCFS-NQ. This trend persisted as workload volumes increased, underscoring the energy-efficient nature of MECT and MEET, which prioritize tasks judiciously based on completion and execution time estimates.

\subsubsection{Wasted Energy Consideration}

\begin{figure}[h]
    \centering
    \includegraphics[width=.45\textwidth]{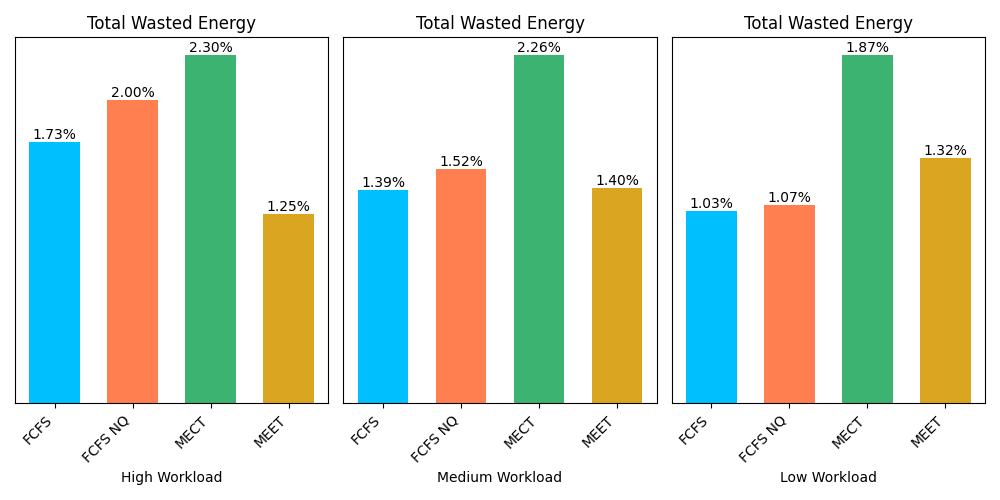}
    \caption{Total Wasted Energy.}
\label{fig:TWE}
\vspace{-.3 cm}
\end{figure}

Additionally, our analysis encompassed an evaluation of wasted energy, reflecting inefficiencies in resource utilization. Notably, MECT and MEET exhibited lower levels of wasted energy across all workload scenarios compared to FCFS and FCFS-NQ. In low workload scenarios, MECT and MEET demonstrated a modest increase in wasted energy, attributable to their proactive task prioritization approach. However, this increase was outweighed by the significant reduction in total consumed energy, resulting in a net gain in energy efficiency. The result is shown in figure \ref{fig:TWE}.   

\subsubsection{Energy Per Completion Analysis}

The assessment also considered energy per completion, a metric indicative of the energy expended per task completed. Here again, MECT and MEET outperformed FCFS and FCFS-NQ across all workload scenarios, exhibiting substantially lower energy per completion values. Particularly in high workload scenarios, where computational demands are heightened, MECT and MEET showcased remarkable efficiency, with energy per completion values of 2.27\% and 0.22\% respectively, highlighting their suitability for resource-constrained environments with demanding task deadlines. The result is shown in figure \ref{fig:EPC}   

These findings underscore the significance of energy-efficient task scheduling algorithms, such as MECT and MEET, in optimizing resource utilization and mitigating energy consumption across diverse workload scenarios. By prioritizing tasks based on comprehensive completion and execution time estimates, these algorithms offer a sustainable and efficient approach to task scheduling in heterogeneous computing environments.

\begin{figure}[h]
    \centering
    \includegraphics[width=.5\textwidth]{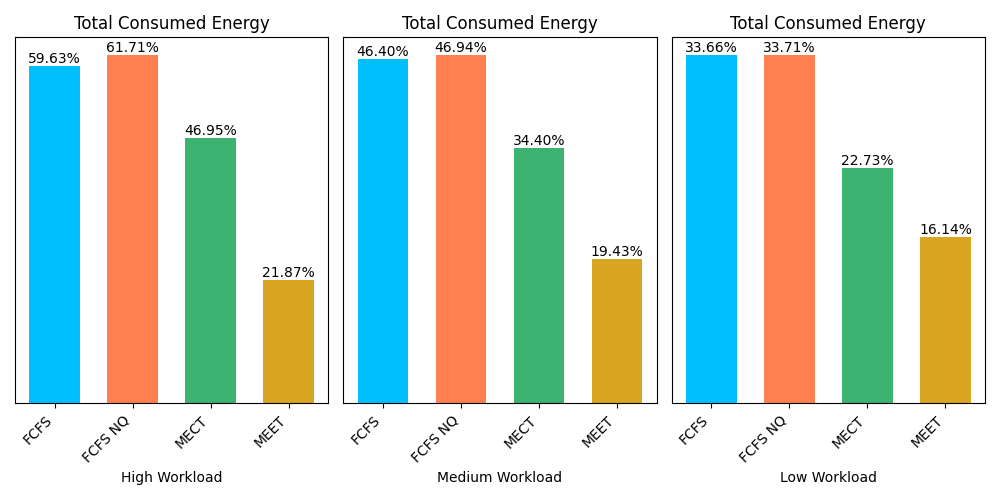}
    \caption{Total Consumed Energy.}
\label{fig:TCE}
\vspace{-.3 cm}
\end{figure}

\begin{figure}[h]
    \centering
    \includegraphics[width=.45\textwidth]{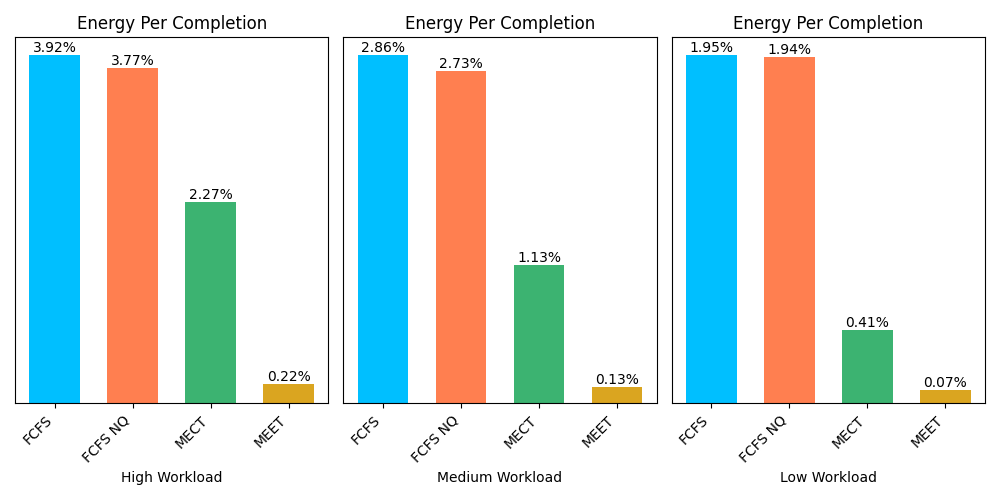}
    \caption{Energy Per Completion.}
\label{fig:EPC}
\vspace{-.3 cm}
\end{figure}

\begin{figure}[h]
    \centering
    \includegraphics[width=.45\textwidth]{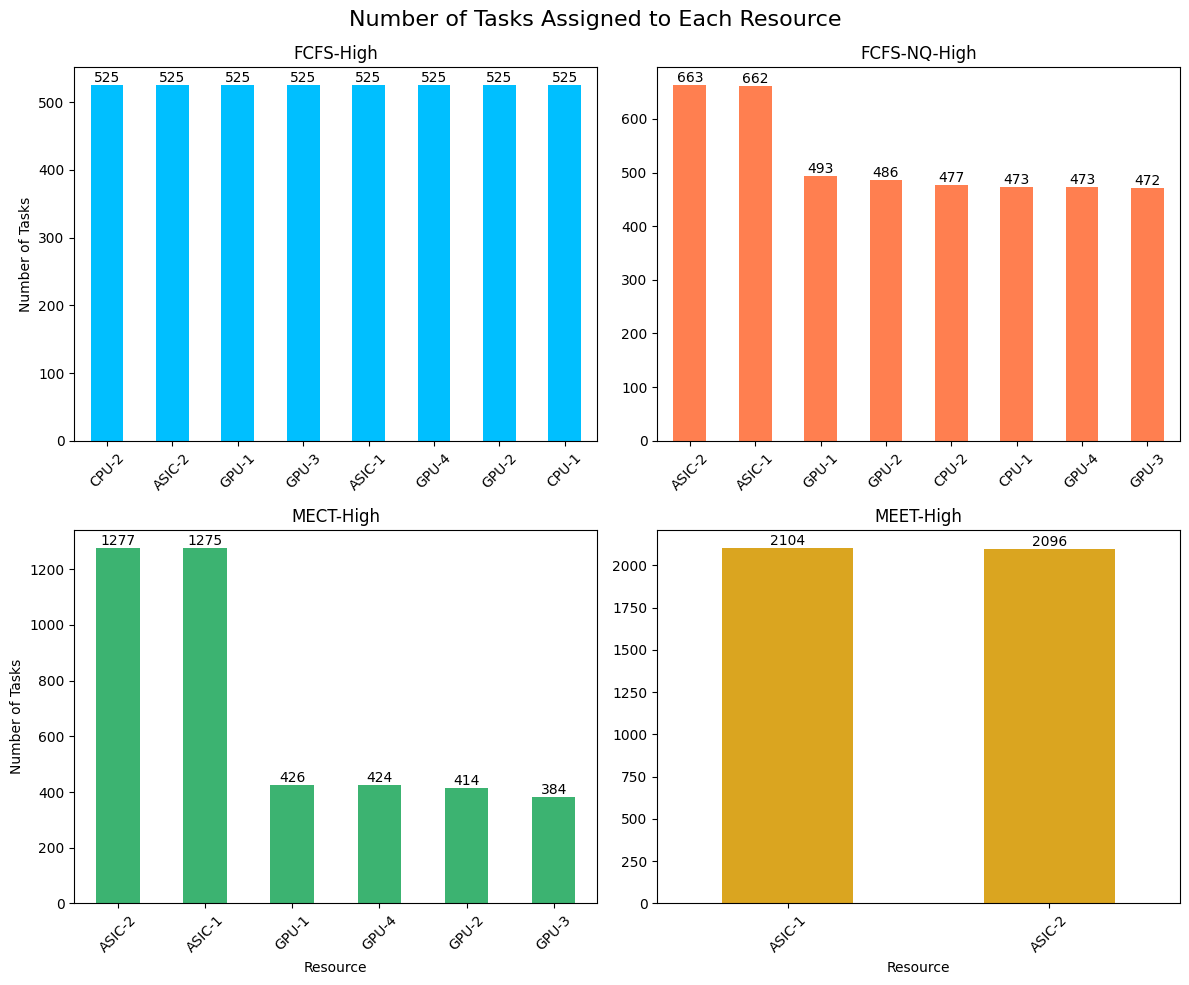}
    \caption{Number of Tasks Assigned to Each Resource.}
\label{fig:overview}
\vspace{-.3 cm}
\end{figure}

\section{Conclusion and Future Works}
In the realm of heterogeneous computing environments, the selection of appropriate task scheduling algorithms plays a pivotal role in optimizing resource utilization, minimizing task completion times, and mitigating energy consumption. Through our comprehensive benchmarking analysis, we have gained valuable insights into the performance characteristics of four distinct scheduling algorithms—First-Come, First-Served (FCFS), FCFS with No Queuing (FCFS-NQ), Minimum Expected Completion Time (MECT), and Minimum Expected Execution Time (MEET)—across varying workload scenarios.

Our findings underscore the nuanced interplay between algorithmic sophistication and workload dynamics. In low workload scenarios, the simplicity of FCFS and FCFS-NQ facilitates efficient task execution, albeit with limitations becoming apparent as workload volumes increase. Conversely, MECT and MEET emerge as robust contenders, dynamically prioritizing tasks based on comprehensive estimates of completion and execution times, thereby optimizing resource utilization and minimizing task completion times across all workload scenarios.

Furthermore, our analysis highlights the critical importance of energy efficiency in task scheduling algorithms. MECT and MEET consistently demonstrate superior energy efficiency compared to FCFS and FCFS-NQ, as evidenced by lower levels of total consumed energy, wasted energy, and energy per completion across all workload scenarios. This underscores their suitability for resource-constrained environments and their potential to significantly reduce operational costs and environmental impact.

In conclusion, our study provides valuable insights into the efficacy of task scheduling algorithms in heterogeneous computing environments. By understanding the unique performance characteristics of each algorithm and their suitability for different workload scenarios, system architects and developers can make informed decisions to optimize resource allocation, minimize task completion times, and enhance energy efficiency in heterogeneous computing environments. Moving forward, further research into algorithm refinement and adaptation to evolving workload dynamics will be instrumental in advancing the efficiency and sustainability of task scheduling in heterogeneous computing environments.

\bibliographystyle{IEEEtran}
\bibliography{cite}
\end{document}